\documentclass[conference]{IEEEtran}
\IEEEoverridecommandlockouts
\usepackage{cite}
\usepackage{amsmath,amssymb,amsfonts}
\usepackage{algorithmic}
\usepackage{graphicx}
\usepackage{textcomp}
\usepackage{xcolor}
\def\BibTeX{{\rm B\kern-.05em{\sc i\kern-.025em b}\kern-.08em
    T\kern-.1667em\lower.7ex\hbox{E}\kern-.125emX}}

\usepackage{todonotes}
\usepackage{tabularx}
\usepackage[hidelinks]{hyperref}
\usepackage{pgfplots}
\usepackage{dblfloatfix}
\pgfplotsset{compat=1.18}
\usepgfplotslibrary{statistics}
\usepackage{tikz}
\usetikzlibrary{patterns}
\usepackage{caption}
\usepackage{multirow}
\usepackage[framemethod=tikz]{mdframed}
\mdfdefinestyle{theoremstyle}{
    linecolor = red,
}
\mdtheorem[style=theoremstyle]{result}{Result}

\usepackage{amssymb}

\definecolor{link}{RGB}{0, 123, 255}

\begin{document}

\title{SciKGDash: The Scientific Knowledge Graph Dashboard for Supporting Knowledge Curation

\thanks{The work was partially funded by the German Research Foundation (DFG) 460234259 (NFDI4DataScience), 501865131 (NFDI4Energy), and 442146713 (NFDI4Ing) within the German National Research Data Infrastructure (NFDI).}
}
\author{
    \IEEEauthorblockN{
    Lena John\IEEEauthorrefmark{1},
    Sören Auer\IEEEauthorrefmark{1},
    Oliver Karras\IEEEauthorrefmark{1}}
    \IEEEauthorblockA{\IEEEauthorrefmark{1}TIB - Leibniz Information Centre for Science and Technology, Hannover, Germany
    \\\{lena.john, soeren.auer, oliver.karras\}@tib.eu}
}

\maketitle

\begin{abstract}
Research knowledge graphs (RKGs) have emerged as essential technology for organizing scientific knowledge, but their success depends heavily on the quality of their underlying content. Knowledge curation is a critical task to ensure the quality of (research) knowledge graphs ((R)KGs), with human curation being the gold standard despite its time- and resource-intensive nature. Automated methods, while efficient, lack the precision of human expertise. Hybrid approaches, combining automated processes with human oversight, offer a promising solution to this challenge. Dashboards can act as supportive tools in hybrid curation approaches, offering real-time updates and visual overviews. This paper presents an action research study, conducted in collaboration with the Curation and Community Building (C\&CB) team of the Open Research Knowledge Graph (ORKG), to explore the development of a dashboard, called SciKGDash, designed to support knowledge curation of the ORKG. SciKGDash serves as a minimum viable product (MVP) tailored to the needs of the C\&CB team, with potential for adaptation to other (R)KGs. An experiment with 15 participants demonstrated the usability of SciKGDash, with successful completion of 4 out of 5 curation tasks in under 5 minutes. In addition, SciKGDash received a positive user experience rating (UEQ score of 1.93). While the tailored solution proved effective for the ORKG, the research also highlights limitations in applying specific quality metrics across diverse (R)KGs. Future work should focus on identifying common quality metrics and enhancing SciKGDash with user-friendly features for querying customized quality metrics. Overall, knowledge curation in RKGs remains an under-explored field, warranting further research.
\end{abstract}


\section{Introduction}
Research knowledge graphs (RKGs) established as a promising technology for organizing scientific knowledge, especially when combined with generative artificial intelligence to develop innovative digital services~\cite{Stocker.2022, Pan.2024}. However, the success of these services depends on the quality of the underlying RKGs~\cite{Chen.2019,Huaman.2021}. For this reason, knowledge curation became a crucial task for ensuring high-quality (research) knowledge graphs ((R)KGs)\footnote{We use RKG when referring specifically to research knowledge graphs and (R)KG when referring to research knowledge graphs and knowledge graphs.}~\cite{Fensel.2020}. 


There are several approaches for supporting knowledge curation, including human-based, automated, and hybrid approaches~\cite{Xue.2022}. Human-based curation ensures high accuracy, but is cumbersome and time-consuming. In contrast, automated approaches can handle large volumes of data efficiently, but often lack precision. Hybrid approaches attempt to balance these trade-offs. While human-based curation is still the gold standard in knowledge curation, hybrid approaches, especially those employing human-in-the-loop (HITL), gain traction~\cite{Peng.2023, Xue.2022}. These approaches leverage the strengths of both humans and machines by integrating human expertise into automated processes. While machines handle the initial data processing and preliminary curation tasks, humans provide oversight, validation, and refinement. 
The incorporation of dashboards into hybrid approaches can further support knowledge curation of (R)KGs by providing a visual and interactive interface that allows human curators to monitor, analyze, and manage the knowledge more effectively~\cite{Gürdür.2019, Patel.2023}. Based on automated (pre-)processing, dashboards can display real-time updates, highlight areas requiring attention, and offer tools for data manipulation and validation~\cite{Vaidyambath.2019, Debattista.2016, Graux.2020, Lezhnina.2022, Shamsabadi.2024}. These capabilities streamline the curation process and enable more informed decision-making, leading to higher quality (R)KGs.

While dashboards are established tools in industry~\cite{Vilarinho.2018,Gürdür.2019, Josko.2017}, their application to support knowledge curation of (R)KGs is limited~\cite{Verma.2023}. The complexity and peculiarities of knowledge curation often require tailored solutions and methods~\cite{Umutcan.2022, Huaman.2021}. As a result, the use of dashboards to support knowledge curation of (R)KGs is under-explored, but shows potential for innovative research~\cite{Verma.2023, Huaman.2021}.

In this paper, we explore the use of dashboards for supporting knowledge curation of RKGs. Given the complexity and peculiarities of knowledge curation, we framed our research approach based on the design science paradigm~\cite{Runeson.2020} by using action research for the solution design and an experiment for the validation. We conducted the action research in close collaboration with the Curation \& Community Building (C\&CB) team of the 
\href{https://orkg.org/}{\textit{\textcolor{link}{Open Research Knowledge Graph (ORKG)}}}~\cite{Auer.2025}. While the entire ORKG team supports the knowledge curation of the ORKG, the C\&CB team is primarily responsible for it. The C\&CB team identified the need for a dashboard to support their knowledge curation efforts. They provided their need as a use case for our research, which aimed to develop a minimal viable product (MVP) of the dashboard for them. In the experiment, we validated the extent to which the MVP is an effective, efficient, and satisfactory solution for their knowledge curation efforts.

As a result of this research approach, we contribute \href{https://orkg.org/curation-dashboard}{\textit{\textcolor{link}{SciKGDash}}}, a dashboard for supporting knowledge curation of the ORKG. Through this real-world use case, we demonstrate how dashboards can support knowledge curation of RKGs and derive key takeaways for adaption to other use cases.
The experiment with 15 subjects showed that human curators can successfully complete on average 4 out of 5 defined curation tasks in less than 5 minutes each by using SciKGDash, whose overall user experience was rated as excellent. These results indicate that the developed MVP of SciKGDash already represents to a large extent an effective, efficient, and satisfactory solution regarding the needs of the C\&CB and entire ORKG team.

Structure: Section~\ref{sec:related_work} reviews related work. Section~\ref{sec:method} describes the research approach. Section~\ref{sec:results} presents the results. Section~\ref{sec:threats_limitations} discusses threats to validity. Section~\ref{sec:discussion} discusses the findings, and Section~\ref{sec:conclusion} concludes the paper.

\section{Related Work}\label{sec:related_work}
In general, curation involves refining, correcting, and completing (R)KGs to improve data quality~\cite{Paulheim.2016, Fensel.2020}, ensuring they remain accurate and usable. 
One of the primary challenges in knowledge curation of RKGs is ensuring high data quality. Quality of (R)KGs is mainly understood as \textit{fit-for-use}/\textit{fit-for-purpose}~\cite{Chen.2019}, which means their quality depends on specific requirements and expectations of users in a particular context and is therefore highly context- and user-dependent. Huaman et al.~\cite{Huaman.2022}, Dorsch et al.~\cite{Dorsch.2023}, and Hogan et al.~\cite{Hogan.2022} provide foundational guidelines and metrics for assessing (R)KGs across multiple dimensions such as accuracy, coverage, coherency, succinctness, and accessibility.

A recurring theme in the knowledge curation of (R)KGs is the use of human-in-the-loop (HITL) methods. While some upstream methods (e.g., \cite{John_Scimantify.2025, John_ExtracTable.2025}) focus on producing 
semi-structured knowledge that is ready for (R)KG import, our concern is the subsequent curation phase. 
Existing HITL approaches mainly incorporate human oversight during the construction and refinement of (R)KGs.
For example, Schatz et al.~\cite{Schatz.2022} propose \textit{KG-CURATE}, where human experts assist in converting plain text into a structured graph. Manzoor et al.~\cite{Manzoor.2023} propose a user-in-the-loop framework where human experts verify predicted parent nodes for new concepts.

Building on the theme of human involvement, Bikaun et al.~\cite{Bikaun.2024} focus on the refinement of existing (R)KGs with their \textit{CleanGraph} tool, which provides an interactive graph view for users to enhance data quality. However, their approach lacks integrated mechanisms for calculating specific quality metrics, leaving users without built-in tools to assess data quality characteristics.
In contrast, Rahman et al.~\cite{Rahman.2024} introduce \textit{Kyurem}, which allows users to program their own widgets for customized graph exploration and quality checks. The widgets facilitate visualization and provide valuable context to support decision-making throughout the workflow. However, users must possess programming skills to fully utilize the system as the interaction occurs in computational notebooks, limiting accessibility to non-technical users.

A more user-friendly alternative is the use of dashboards, which have been widely adopted across various domains to streamline quality monitoring in real-time without requiring advanced technical skills. For example, Gürdür et al.~\cite{Gürdür.2019} propose a methodology for linked enterprise data quality assessment through information visualizations, which could be adapted for (R)KG environments. Debattista et al.~\cite{Debattista.2016} present \textit{Luzzu}, a customizable framework for data consumers to assess data quality in linked datasets and thus find datasets that \textit{fit their purpose}. Vaidyambath et al.~\cite{Vaidyambath.2019} enhance the \textit{Luzzu} framework by adding an analytics dashboard that suggests and intelligently prioritizes data quality issues, while also enabling root cause analysis (RCA) through RDF graph generation to assist in diagnosing these issues. Graux et al.~\cite{Graux.2020} focus on real-time monitoring by developing a Wikipedia dashboard that highlights the most recent edits and active users, offering an overview of ongoing changes within the platform. Pellegrino et al.~\cite{Pellegrino.2024} present \href{http://www.isislab.it:12280/kgheartbeat/kgheartbeat}{\textit{\textcolor{link}{KGHeartBeat}}}, an online dashboard that assesses and compares the quality of publicly available (R)KGs based on data-level metrics and metadata attributes. However, its generic framework does not allow for an in-depth analysis of highly heterogeneous or domain-specific graphs. To the best of our knowledge, this work is most closely related to our presented research.

In the context of the ORKG, Lezhina et al.~\cite{Lezhnina.2022} present a dashboard that aims to improve the user interaction of data consumers with research data through a visual interface. Similarly, Shamsabadi et al.~\cite{Shamsabadi.2024} also present a \href{https://orkg.org/usecases/r0-estimates}{\textit{\textcolor{link}{dashboard}}} for the ORKG that allows data consumers to visualize the results of LLM-based information extraction of $R_0$ estimates. While these dashboards facilitate exploration and understanding of scientific content for data consumers, they primarily focus on presentation rather than active knowledge curation. Thus, despite the potential benefits of dashboards in enhancing data access for data consumers, their specific application to knowledge curation in the dynamic research environment of the ORKG remains limited and under-explored.\vspace{0.1cm}

\begin{figure*}[!b]
    \centering
    \includegraphics[width=0.6\textwidth]{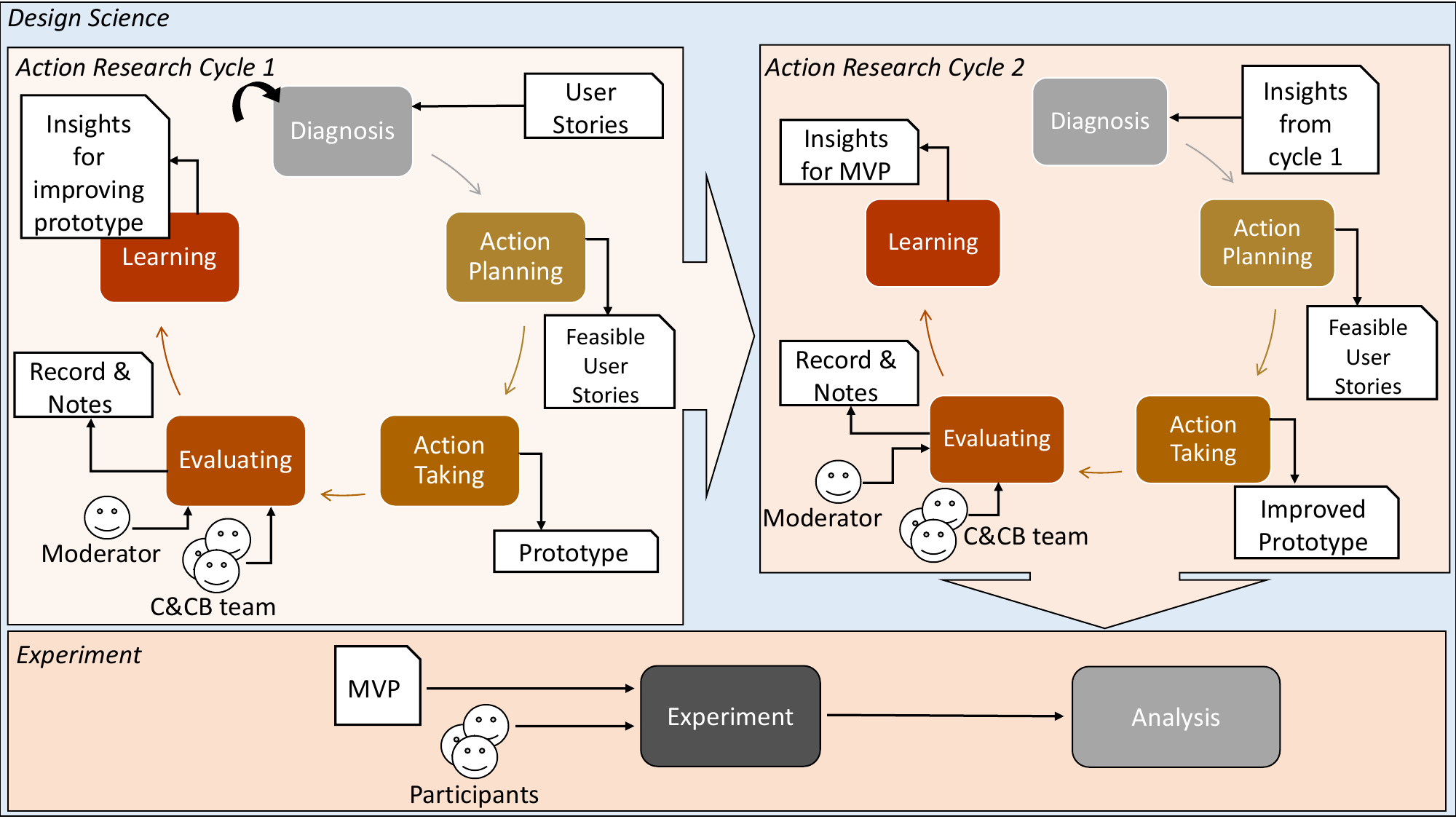}
    \caption{Design science paradigm with two action research cycles for the solution design and an experiment for validation.}
    \label{fig:actionresearch}
\end{figure*}

The presented related work underscores that knowledge curation of (R)KGs often focuses on enabling individual users to assess specific quality characteristics of their (R)KGs. While this approach is suitable for smaller, user-maintained (R)KGs, it raises the question of how to support knowledge curation of larger, community-driven (R)KGs, such as DBpedia or ORKG. Typically, communities address issues reactively as they arise, leading to the challenge of consistent issue resolution across entire (R)KGs. A systematic approach is needed to detect and address recurring issues at scale for effective knowledge curation of (R)KGs.

For this reason, we propose incorporating dashboards into hybrid approaches for knowledge curation of RKGs. By starting with the global view, human curators can gain deeper insights into how and where quality issues, reflected in the measured Key Performance Indicators (KPIs), manifest across the entire RKG. This approach provides human curators with an improved overview and facilitates \mbox{(semi-)}automated refactoring at scale to resolve systematic issues in the entire RKG.
A key challenge of our proposed approach lies in the subjective nature of RKG quality, which is often understood as \textit{fit-for-use}/\textit{fit-for-purpose}, making it context- and user-dependent, and generic solutions less beneficial. For this reason, we developed our approach based on a specific RKG, the ORKG, and in close collaboration with its C\&CB team. This decision is particularly important to develop a solution that provides practical added value and is therefore also used in the long term. According to Pellegrino et al.~\cite{Pellegrino.2024}, there is a lack of accessible and actively maintained tools for knowledge curation and RKG quality assessment. We reflect on the lessons learned during the development of SciKGDash and generalize them into key takeaways of our approach. These insights aim to guide other researchers and human curators in creating dashboards for their own (R)KGs, aligned with their particular quality needs.

\section{Method}\label{sec:method}
We adopt the design science paradigm~\cite{Runeson.2020} to develop the MVP of SciKGDash through two iterative cycles of action research~\cite{Staron.2020} (see \figurename{~\ref{fig:actionresearch}}). The MVP in our context is understood as a ``\textit{set of minimal requirements, which meet the needs of the core group of early adopters or users}''~\cite{Lenarduzzi.2016}. The core group consists of the C\&CB team and ORKG team members who curate knowledge in the ORKG.

We employ design science as it provides a structured framework to develop and evaluate a solution design for the MVP, ensuring the MVP meets the objectives. Action research is selected as it allows for interventions within the organizational context of the ORKG, making the resulting solution design directly applicable and usable by the C\&CB team.

\subsection{Action Research for Solution Design}
A close collaboration between the organizational partner, the C\&CB team of the ORKG, and the authors of this paper is essential for action research, as it enables targeted interventions in curation processes while simultaneously generating academic insights. By integrating the insights into the iterative development of the MVP, the solution design remains aligned with the C\&CB team’s needs and adaptable to the dynamic nature of knowledge curation.

Over six months, we gained access to the C\&CB team's communication channels and obtained valuable insights into their daily work through an observational role. The C\&CB team also allowed us to attend their weekly meetings. 
Curators focus on assessing domain-specific knowledge and data quality, whereas community-building experts examine modeling practices and user behavior to improve support.

The process of action research began with the diagnosing 
phase, where \href{https://zenodo.org/records/16908464/files/User_Stories.csv?download=1}{\textit{\textcolor{link}{user stories}}} were provided by experts in the C\&CB team. In each cycle, the most urgent and feasible user stories were prioritized, focusing on depth rather than breadth by implementing selected features comprehensively (see \tablename{~\ref{tab:user_stories}}). Over both cycles, the prototype evolved by expanding its functionality and refining existing features. The resulting prototype from each cycle was evaluated in a two-month interval through online sessions using pluralistic walkthrough~\cite{Hollingsed.2007}, engaging the entire C\&CB team with eight members. A moderator, familiar with both the author team and the C\&CB team's workflows, guided the sessions and provided additional context during the evaluations. The sessions were recorded for detailed post-analysis. We presented the current prototype and demonstrated how the dashboard could be used for various tasks. Afterwards, feedback was collected regarding the dashboard, focusing on its usability and whether it aligned with the C\&CB team's envisioned workflows for curation and community-building tasks.

A key design consideration was that the dashboard should not directly modify any content of the ORKG but instead serve as a supportive tool to enhance the C\&CB team's workflows by leveraging machine assistance. Continuous feedback from the C\&CB team was provided informally through a Skype group, where we consulted the C\&CB team for input on critical decisions, such as privacy handling of user data and data visualization choices. This iterative approach facilitated rapid adjustments and contributed to faster implementation.

After completing both cycles, the final MVP emerged as \href{https://orkg.org/curation-dashboard}{\textit{\textcolor{link}{SciKGDash}}}, which was subsequently analyzed in a controlled experiment to assess its usability in terms of effectiveness, efficiency, and satisfaction. By combining design science and action research, it is ensured that the developed MVP is not only theoretically grounded but also practically relevant and applicable in a real-world scenario.

\subsection{Experiment for Validation}
With the goal definition template~\cite{Wohlin.2012}, we specify the goal for the experiment to validate the MVP of SciKGDash:

\begin{mdframed}
    \hypertarget{ResearchGoal}{\textit{Research goal:}} \textbf{Analyze} the developed MVP
    \textbf{for the purpose of} supporting the process of curation and community building of RKGs 
    \textbf{with respect to} usability (effectivenss, efficiency, and satisfaction)
    \textbf{from the point of view of} the ORKG team, especially the C\&CB team 
    \textbf{in the context of} the ORKG.
\end{mdframed}
\vspace{-0.2cm}

Based on this goal, we ask the following research question:

\begin{mdframed}
    \hypertarget{ResearchQuestion}{\textit{Research question:}} How well does the developed MVP support the ORKG team, especially the C\&CB team, regarding their specified needs concerning usability in the curation and community building of the ORKG?
\end{mdframed}
\vspace{-0.2cm}

Usability is defined as the combination of the factors: effectiveness, efficiency and satisfaction~\cite{Lawal.2022}. Effectiveness and efficiency is measured using objective and subjective metrics, while satisfaction will be assessed using the short user experience questionnaire (\href{https://www.ueq-online.org/}{\textit{\textcolor{link}{UEQ}}})~\cite{Schrepp.2017}. The participants also subjectively evaluate how well the dashboard meets the team's needs. In \tablename{~\ref{tab:variable_experiment}}, we define the dependent and independent variables of the experiment:

\begin{footnotesize}
\begin{table}[htbp]
   \vspace{-2mm}
    \caption{Dependent and independent variables.}
    \begin{tabularx}{\linewidth}{|p{1.3cm}|p{2.7cm}|X|}
        \hline
         \multicolumn{1}{|c}{Type} & \multicolumn{1}{|c|}{Variable} & Explanation \\ \hline\hline
         Independent & SciKGDash & The provided MVP. \\ \hline
         \multirow{12}{*}{Dependent} & \multirow{3}{*}{Objective effectiveness} & A boolean indicating if the participant successfully completed a task or not.\\ \cline{2-3}
         & \multirow{3}{*}{Subjective effectiveness} & Agreement with ``\textit{I finished the task successfully.}'' on a Likert scale.\\\cline{2-3}
         & \multirow{2}{*}{Objective efficiency} & The measured time a participant needs to complete a task.\\\cline{2-3}
         & \multirow{2}{*}{Subjective efficiency} & Agreement with ``\textit{I finished the task fast.}'' on a Likert scale.\\\cline{2-3}
         & \multirow{2}{*}{Objective satisfaction} & The short UEQ score after completing all tasks.\\\cline{2-3}
         & \multirow{3}{*}{Subjective satisfaction} & Agreement with the fulfillment of five needs in the MVP on a Likert scale.\\\hline
    \end{tabularx}
    \label{tab:variable_experiment}
    \vspace{-2mm}
\end{table}
\end{footnotesize}

\subsubsection{Setting} 
The experiment involved participants from the ORKG team, including members of the C\&CB team, recruited through convenience sampling via an open call, with voluntary participation. A within-subjects design was used without a control group, as the independent variable had only one treatment condition. Sessions lasted 30 minutes and were conducted in person or online to accommodate the participants' geographic distribution. Participants received \href{https://doi.org/10.5281/zenodo.16908463}{\textit{\textcolor{link}{textual materials}}} containing five tasks (see~\tablename{~\ref{tab:tasks}}) and evaluation sections, complemented by oral support if needed. The tasks focused on specific curation and community-building activities of the C\&CB team, ensuring a comprehensive evaluation.

\begin{footnotesize}
\begin{table}[htbp]
    \vspace{-2mm}
    \caption{Tasks for the experiment.}
    \begin{tabularx}{0.5\textwidth}{|l|X|}\hline
    \multirow{2}{*}{Task 1} & Identify the Predicate with the least amount of duplicates which is lacking a description, open the respective pages in the ORKG.\\\hline
    \multirow{3}{*}{Task 2.1} & Fetch visitor data for a specific date range, find the most frequently used path (edge) between pages, and identify where users most often go next from the end node of that path.\\\hline
    \multirow{2}{*}{Task 2.2} & Identify the most used path of minimum length 3. Determine the number of occurences.\\\hline
    \multirow{2}{*}{Task 3} & Identify the Paper with the least amount of Statements and open it in the ORKG.\\\hline
    \multirow{3}{*}{Task 4} & Comment on the Paper with the fewest Statements, choosing the most appropriate comment type and providing a brief explanation.\\\hline
    \end{tabularx}
    \label{tab:tasks}
    \vspace{-2mm}
\end{table}
\end{footnotesize}

Online participants had to switch between applications to complete the tasks and fill out the materials. Sessions were recorded with participants' prior consent, capturing screen activity and audio to support the think-aloud method~\cite{Someren.1994} for detailed post-session analysis. After completing the five tasks, participants could independently explore the dashboard and provide additional feedback for future development.

\subsubsection{Material}
Participants first provided demographic data and then completed five tasks (cf.~\tablename{~\ref{tab:tasks}}), each designed from the perspective of a curator or community-building expert. Each task included a brief explanation of its purpose to help participants understand its relevance to curation and community-building workflows. Co-developed with the moderator, the tasks were designed to align with curation and community-building workflows and to test the MVP’s capabilities.
All \href{https://doi.org/10.5281/zenodo.16908463}{\textit{\textcolor{link}{materials}}}, including user stories, tasks, and results, are openly accessible.

After each task, participants rated their subjective effectiveness and efficiency using a 5-point Likert scale. We used the think-aloud method to ensure that participants clearly explained the reasons for their ratings. In this way, we mitigated the risk of neutral responses being misinterpreted~\cite{Pornel.2013}. This approach aimed to identify discrepancies between the user task model (how users believe tasks should be performed) and the system task model (how the system assumes tasks should be performed)~\cite{Mori.2002}, allowing for user interface (UI) improvements for users’ desired activities.

The tasks were not randomized as some depended on intermediate outcomes or a learning effect. At the end of the session, participants completed the UEQ and assessed the fulfillment of specified needs (see \tablename{~\ref{tab:needs_of_team}}). These needs were defined by the authors, validated by the moderator, derived from high-priority \href{https://doi.org/10.5281/zenodo.16908463}{\textit{\textcolor{link}{user stories}}} and MVP features, focusing on the core tasks and goals of the C\&CB team’s curation and community-building activities.

\begin{footnotesize}
\begin{table}[htbp]
    \caption{Specified needs of the team.}
    \begin{tabularx}{0.49\textwidth}{|l|X|}\hline
    N1 & The dashboard allows the tracking of visitor paths.\\\hline
    \multirow{2}{*}{N2} & The dashboard makes it possible to indicate if a resource is inaccurate or incomplete.\\\hline
    \multirow{2}{*}{N3} & The dashboard provides tools for monitoring the quality of the ORKG.\\\hline
    N4 & The dashboard provides an overview of areas for improvement.\\\hline
    N5 & The dashboard lists how active the internal and external users are.\\\hline
    \end{tabularx}
    \label{tab:needs_of_team}
    \vspace{-3mm}
\end{table}
\end{footnotesize}

Four out of five needs (N1-N4) were implemented in the MVP, with N5 serving as a control question to assess participant attention. Although initially prioritized by the C\&CB team, N5 could not be implemented due to ORKG backend limitations, which prevent distinguishing internal (ORKG team) and external users. The ORKG tracks minimal information about its users to protect user privacy. The control question was presented last to minimize influence on participants' behavior or subsequent responses~\cite{Kung.2017}.

\subsubsection{Data analysis}
We framed the experiment by formulating hypotheses for effectiveness, efficiency, and satisfaction.\vspace{0.2cm}

Effectiveness is assessed based on aggregated results across all participants. Following Fr\o{}kj\ae{}r et al.~\cite{Frokjaer.2000}, effectiveness is defined as the accuracy and completeness in achieving specific goals. We focus on completeness, whether participants achieve the task goal, regardless of the methods used within the dashboard.

\begin{table}[htbp]
    \vspace{-2mm}
    \centering
    \begin{tabularx}{\linewidth}{lX}
        ($H_{0, Effectiveness}$): & \footnotesize The participants complete maximum 3 out of 5 tasks successfully.\\
        ($H_{1, Effectiveness}$): &\footnotesize The participants complete more than 3 out of 5 tasks successfully.\\
    \end{tabularx}
    \vspace{-3mm}
\end{table}

Efficiency is evaluated based on each participant's measured time to complete a task. An expected duration of 5 minutes per task is set, accounting for both familiarization with the UI and task execution. The dashboard should be lightweight and not overly time-consuming for participants. Considering the task complexity and participants' skill levels, a 5-minute duration per task is deemed appropriate for the hypothesis.

\begin{table}[htbp]
   \vspace{-2mm}
    \centering
    \begin{tabularx}{\linewidth}{lX}
        ($H_{0, Efficiency}$): &\footnotesize The measured time a participant needs to complete a task is 5 minutes or more.\\
        ($H_{1, Efficiency}$): &\footnotesize The measured time a participant needs to complete a task is less than 5 minutes.\\
    \end{tabularx}
    \vspace{-3mm}
\end{table}

Satisfaction is evaluated using the short UEQ. A mean value of $>0.8$ on the UEQ scale indicates a positive user experience.

\begin{table}[htbp]
    \vspace{-2mm}
    \centering
    \begin{tabularx}{\linewidth}{lX}
        ($H_{0, Satisfaction}$): &\footnotesize The participants have no positive user experience with the dashboard (UEQ $<=$ 0.8).\\
        ($H_{1, Satisfaction}$): &\footnotesize The participants have a positive user experience with the dashboard (UEQ $>$ 0.8).\\
    \end{tabularx}
    \vspace{-3mm}
\end{table}

For the subjective measurements, we applied descriptive statistics to summarize and analyze the data. The choice of statistical tests for objective measurements depended on the normality of the data distribution. We assessed the normality using the Shapiro-Wilk test. For data following a normal distribution, a one-sample t-test was employed. Otherwise, the one-sample Wilcoxon signed-rank test was used. These hypothesis tests were conducted to evaluate the significance of the measured indicators and to assess whether the dashboard met the predefined objectives.

\section{Results}\label{sec:results}

\subsection{Action Research Results}
The \href{https://doi.org/10.5281/zenodo.16908463}{\textit{\textcolor{link}{user stories provided}}} by the C\&CB team formed the foundation for developing the MVP. They were prioritized based on their importance, and we categorized them into realizable features and the team's needs, starting with the highest-priority ones. As Cohn~\cite{Cohn.2004} emphasizes, user stories are inherently iterative and not permanent artifacts but evolve throughout development. 
Over time, the user stories were refined and transformed into features, shown in \tablename{~\ref{tab:user_stories}}, classified by dashboard type (strategic, analytical, and operational)~\cite{Few.2005}. According to Few~\cite{Few.2005}, strategic dashboard features support long-term decision-making, analytical dashboard features provide contextual depth and interactivity, and operational dashboard features respond to real-time data needs.

The user stories revealed challenges in the knowledge curation of RKGs, particularly within the context of the ORKG, highlighting the need to make existing data more accessible for content quality assessment. The focus was on developing components that move beyond generic structural metrics toward KPIs tailored to the ORKG's schema and curation workflows. The team currently lacks an overview of such indicators, as their identification requires manual exploration of the graph. Further on, user behavior on the platform is analyzed to derive insights into modeling practices, which inform ongoing interface refinements. 

Given this context, a baseline comparison with \textit{SciKGDash} would be unfair, as the required features do not yet exist in the current system. Likewise, comparison with existing tools are of limited value, since most address generic metrics rather than the specific characteristics of (R)KGs. 

Each feature was developed or refined across one or both cycles, depending on its complexity and emerging insights. The dashboard was initially built in a \href{https://anonymous.4open.science/r/scikgdash-prototype-8E62}{\textit{\textcolor{link}{React-based environment}}} and now migrated to \href{https://gitlab.com/TIBHannover/orkg/scikgdash}{\textit{\textcolor{link}{Next.js}}}. The dashboard \href{https://orkg.org/curation-dashboard}{\textit{\textcolor{link}{SciKGDash is live}}} and under active development.
 
\begin{footnotesize}
\begin{table}[htbp]
    \vspace{-2mm}
    \caption{Implemented MVP features by dashboard type.}
    \begin{tabularx}{\linewidth}{|l|c|X|}
        \hline
         Feature & Cycle & Dashboard type\\\hline\hline
         \#Predicates without a description & 1& Strategic\\\hline
         \#Classes without a description & 1& Strategic\\\hline
         Duplicate Predicates & 1 and 2& Strategic/operational\\\hline
         Unused Resources& 1& Strategic\\\hline
         Unlabeled Resources& 1& Strategic\\\hline
         Map of visitor paths& 2& Analytic\\\hline
         \#Papers in a research field& 2& Operational\\\hline
         \#Statements per Paper& 2& Operational\\\hline
         \#empty cells in a Comparison&1 and 2 & Operational\\\hline
         Overview of Template creation& 1 and 2& Operational\\\hline
         Make comments about resources& 2& Operational\\\hline
    \end{tabularx}
    \label{tab:user_stories}
    \vspace{-2mm}
\end{table}
\end{footnotesize}

\subsubsection{First Cycle}
During the first cycle, the technical setup was established, using a React frontend and Python Flask backend. Data was retrieved from the \href{http://tibhannover.gitlab.io/orkg/orkg-backend/api-doc/}{\textit{\textcolor{link}{ORKG REST API}}}, its SPARQL endpoint, and the Matomo API for visitor statistics. Core features were implemented to address basic curation needs, such as identifying unused resources. During the evaluation, it was decided to aggregate visitor statistics to preserve user anonymity, as ORKG URLs could be linked to individual logged-in users. During the learning phase, the C\&CB team emphasized the importance of a comment functionality, which became a priority in the second cycle. This feature, integrated with a database in the backend, enables the team to track and manage resources that need attention, serving as an internal issue tracker for curation and community building tasks.

\subsubsection{Second Cycle}
The second cycle focused on scalability and modularity to support multiple (R)KGs. Reusable query structures now enable tasks like identifying classes without descriptions, common in (R)KGs such as DBpedia, demonstrating a proof of concept for cross-graph quality assessment.

Additional features included a Matomo visitor network, visualizing page relationships and transition frequencies, and a visitor path list for analyzing sequential page visits. We developed a new ORKG backend endpoint to retrieve statement counts per paper. Generalizing the dashboard proved challenging due to varying conceptual structures across (R)KGs. While the challenges highlight the need for adaptable, modular knowledge curation solutions, they also emphasize the importance of tailored approaches.


\subsection{Experiment Results}
\newcommand{\twert}{$t$}
\newcommand{\pwert}{$p$}
\newcommand{\pwertA}{$p_a$}
\newcommand{\zwert}{$Z$}
Fifteen participants took part in the experiment conducted in April 2024. The participation was voluntary, and seven of the participants are part of the C\&CB team and had been involved in evaluation sessions during at least one action research cycle. 

\subsubsection{Correlation}
Seven participants of the C\&CB team took part in at least one action research evaluation session. We investigated whether the number of evaluation sessions attended influenced each objective variable, as repeated participation could improve performance by increasing familiarity with the dashboard. Such relationships could affect the statistical analysis. Spearman’s rank correlation was used, and the results are summarized in \tablename{~\ref{tab:spearmanStats}}.

\begin{footnotesize}
\begin{table}[!htb]
    \caption{Spearman's correlation coefficients between the number of evaluations participated and each objective variable.}
    \begin{tabularx}{\linewidth}{|X|c|c|}
        \hline
        \multirow{2}{*}{Objective variable} & \multicolumn{2}{c|}{Number of evaluations participated}\\
        \cline{2-3}
         & $\rho$ & \pwert \\
         \hline\hline
         Task 1 Efficiency & $-0,349$ & $0,202$ \\\hline
         Task 2.1 Efficiency & $-0,089$ & $0,752$\\\hline
         Task 2.2 Efficiency & $0,225$ & $0,420$ \\\hline
         Task 3 Efficiency & $-0,024$ & $0,933$ \\\hline
         Task 4 Efficiency & $-0,355$ & $0,195$ \\\hline
         Effectiveness & $-0,0253$ & $0,364$ \\\hline
         UEQ & $0,427$ & $0,112$ \\\hline
    \end{tabularx}
    \label{tab:spearmanStats}
\end{table}
\end{footnotesize}

None of the results is statistically significant ($p > 0.05$), which indicates that there is no correlation between the number of evaluations participated and the measurements of the objective variables. Consequently, the data collected from all participants can be analyzed together as one group.

\subsubsection{Effectiveness}
\begin{figure}[htbp]
    \centering
    \footnotesize
    \vspace{-0.2cm}
    \begin{tikzpicture}
        \begin{axis}[
            xbar stacked, 
            ytick=data, 
            yticklabels={T1, Task 2.1, Task 2.2, Task 3, Task 4}, 
            bar width=0.3cm,
            width=0.47\textwidth,
            height=3.5cm,
            xlabel={Number of responses}, 
            symbolic y coords={T1, T2.1, T2.2, T3, T4},
            nodes near coords, 
            legend style={at={(0.5,1.1)},anchor=south,legend columns=-1}, 
            xmin=0, xmax=15,
            xtick={0,1,...,15},
            enlarge y limits=0.2, 
            yticklabel style={align=center},
            grid=both,
        ]
            \addplot[fill=red,fill opacity=0.5, text opacity=1] coordinates {(0,T1) (0,T2.1) (0,T2.2) (0,T3) (0,T4)};
            \addplot[fill=orange,fill opacity=0.5, text opacity=1] coordinates {(0,T1) (1,T2.1) (0,T2.2) (0,T3) (1,T4)};
            \addplot[fill=yellow,fill opacity=0.5, text opacity=1] coordinates {(0,T1) (2,T2.1) (1,T2.2) (0,T3) (0,T4)};
            \addplot[fill=teal,fill opacity=0.5, text opacity=1] coordinates {(2,T1) (2,T2.1) (2,T2.2) (1,T3) (0,T4)};
            \addplot[fill=green,fill opacity=0.5, text opacity=1] coordinates {(13,T1) (10,T2.1) (12,T2.2) (14,T3) (14,T4)};
            \legend{Strongly disagree, Disagree, Neutral, Agree, Strongly agree}
        \end{axis}
        \node[anchor=east] at (7.9,0.25) {$\tilde{x} = 5$};
        \node[anchor=east] at (7.9,0.6) {$\tilde{x} = 5$};
        \node[anchor=east] at (7.9,0.95) {$\tilde{x} = 5$};
        \node[anchor=east] at (7.9,1.3) {$\tilde{x} = 5$};
        \node[anchor=east] at (7.9,1.65) {$\tilde{x} = 5$};
    \end{tikzpicture}
    \caption{Agreement with: ``I finished the task successfully.''.}
    \label{fig:effectivenessLikert}
    \vspace{-0.3cm}
\end{figure}
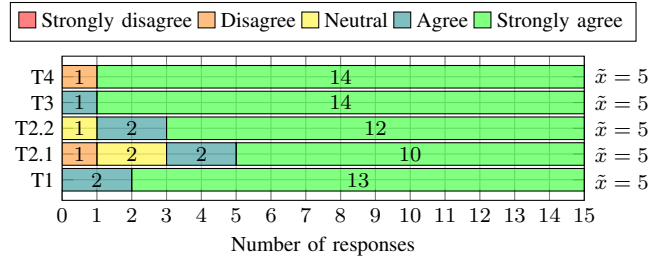

\figurename{~\ref{fig:effectivenessLikert}} shows that the majority of participants rated their task completion for each task as successful, with the median response for all tasks of ``strongly agree'' ($\tilde{x} = 5$). Task~2.1 shows the largest variation in responses, indicating potential complexity. Overall, the feedback was highly positive, with small deviations.

\begin{figure}[!h]
    \centering
    \footnotesize
    \vspace{-0.2cm}
    \begin{tikzpicture}
        \begin{axis}[
            xbar stacked, 
            ytick=data, 
            yticklabels={T1, Task 2.1, Task 2.2, Task 3, Task 4}, 
            bar width=0.3cm,
            width=0.47\textwidth,
            height=3.5cm,
            xlabel={Number of responses}, 
            symbolic y coords={T1, T2.1, T2.2, T3, T4},
            nodes near coords, 
            legend style={at={(0.5,1.1)},anchor=south,legend columns=-1}, 
            xmin=0, xmax=15,
            xtick={0,1,...,15},
            enlarge y limits=0.2, 
            yticklabel style={align=center},
            grid=both,
        ]
            \addplot[fill=red, fill opacity=0.4, text opacity=1] coordinates {(2,T1) (1,T2.1) (4,T2.2) (4,T3) (2,T4)};
            \addplot[fill=green, fill opacity=0.4, text opacity=1] coordinates {(13,T1) (14,T2.1) (11,T2.2) (11,T3) (13,T4)};
            
            \legend{Not successful, Successful};
        \end{axis}
    \end{tikzpicture}
    \caption{Task completion outcomes per task.}
    \label{fig:effectivenesPerTask}
    \vspace{-0.3cm}
\end{figure}
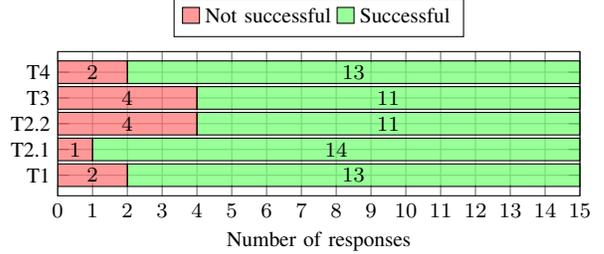

Regarding the objective effectiveness, \figurename{~\ref{fig:effectivenesPerTask}} shows that the participants successfully completed the majority of tasks. Tasks~2.2 and 3 had the highest number of unsuccessful completions, with four participants failing each task. Tasks~1, 2.1, and 4 had only one respectively two unsuccessful completions.

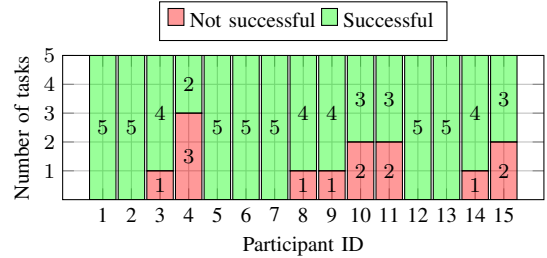
\begin{figure}[htbp]
    \centering
    \footnotesize
    \vspace{-2mm}
    \begin{tikzpicture}
        \begin{axis}[
            ybar stacked,
            xtick=data,
            width=0.47\textwidth,
            xticklabels={1,2,3,4,5,6,7,8,9,10,11,12,13,14,15},
            width=0.44\textwidth,
            height=3.5cm,
            ylabel={Number of tasks},
            xlabel={Participant ID},
            legend style={at={(0.5,1.1)},anchor=south,legend columns=-1},  
            ytick={1,2,3,4,5},
            ymin=0, ymax=5,
            grid,
            nodes near coords,
        ]
            
            \addplot[fill=red, fill opacity=0.4, text opacity=1] coordinates {(1,0) (2,0) (3,1) (4,3) (5,0) (6,0) (7,0) (8,1) (9,1) (10,2) (11,2) (12,0) (13,0) (14,1) (15,2)};
            \addplot[fill=green, fill opacity=0.4, text opacity=1] coordinates {(1,5) (2,5) (3,4) (4,2) (5,5) (6,5) (7,5) (8,4) (9,4) (10,3) (11,3) (12,5) (13,5) (14,4) (15,3)};
            \legend{Not successful, Successful};
        \end{axis}
    \end{tikzpicture}
    \caption{Task completion outcomes per participant.}
    \label{fig:effectivenessPerPerson}
    \vspace{-3mm}
\end{figure}

\figurename{~\ref{fig:effectivenessPerPerson}} illustrates the success rate per participant. Eleven participants successfully completed at least four out of five tasks, three participants completed three tasks, and one participant completed only two tasks. These results indicate an intuitive UI that enabled most participants to accomplish the tasks successfully.

The data from \figurename{~\ref{fig:effectivenessPerPerson}} was statistically tested. The test results in \tablename{~\ref{tab:effectivenessStatistics}} show a significant difference between the measured effectiveness ($Mdn=4$, $n=15$) and the expected effectivness ($Mdn=3$, $n=15$), ($Z=2.87$, $p=0.004$, $r=0.8$). Therefore, the null hypothesis $H_{0, Effectiveness}$ is rejected, indicating that participants completed more than three tasks successfully.

\begin{mdframed}
     \hypertarget{R1}{\textbf{Result 1: Effectiveness.}} The participants successfully completed more than half of the tasks.
\end{mdframed}
\vspace{-0.2cm}

\begin{table}[htbp]
    \begin{footnotesize}
    \vspace{-2mm}
    \caption{Results of statistical test regarding effectiveness.}
    \begin{tabularx}{0.49\textwidth}{|X|c|c|c|c|}
        \hline
        \multicolumn{1}{|l|}{\multirow{2}{*}{Effectiveness}} & \multicolumn{2}{c|}{Shapiro-Wilk} & \multicolumn{2}{c|}{Wilcoxon signed-rank} \\
        \cline{2-5}
         & $W$ & $p$ & \zwert & \pwert \\
        \hline\hline
        Aggregated task completions (T1 to T4) & \multirow{2}{*}{$0.816$} & \multirow{2}{*}{$0.00594$} & \multirow{2}{*}{$2.87$} & \multirow{2}{*}{$0.004$} \\
        \hline
    \end{tabularx}
    \label{tab:effectivenessStatistics}
    \end{footnotesize}
    \vspace{-3mm}
\end{table}

\subsubsection{Efficiency}
\begin{figure}[htbp]
    \centering
    \footnotesize
    \vspace{-2mm}
    \begin{tikzpicture}
        \begin{axis}[
            xbar stacked, 
            ytick=data, 
            yticklabels={T1, T2.1, T2.2, T3, T4}, 
            bar width=0.3cm,
            width=0.47\textwidth,
            height=3.5cm,
            xlabel={Number of responses}, 
            symbolic y coords={T1, T2.1, T2.2, T3, T4},
            nodes near coords, 
            legend style={at={(0.5,1.1)},anchor=south,legend columns=-1}, 
            xmin=0, xmax=15,
            xtick={0,1,...,15},
            enlarge y limits=0.2, 
            yticklabel style={align=center},
            grid=both,
        ]
            \addplot[fill=red,fill opacity=0.5, text opacity=1] coordinates {(0,T1) (1,T2.1) (1,T2.2) (0,T3) (0,T4)};
            \addplot[fill=orange,fill opacity=0.5, text opacity=1] coordinates {(0,T1) (1,T2.1) (3,T2.2) (0,T3) (0,T4)};
            \addplot[fill=yellow,fill opacity=0.5, text opacity=1] coordinates {(3,T1) (3,T2.1) (0,T2.2) (1,T3) (1,T4)};
            \addplot[fill=teal,fill opacity=0.5, text opacity=1] coordinates {(5,T1) (9,T2.1) (5,T2.2) (3,T3) (4,T4)};
            \addplot[fill=green,fill opacity=0.5, text opacity=1] coordinates {(7,T1) (1,T2.1) (6,T2.2) (11,T3) (10,T4)};
            \legend{Strongly disagree, Disagree, Neutral, Agree, Strongly agree}
        \end{axis}
        \node[anchor=east] at (7.9,0.25) {$\tilde{x} = 4$};
        \node[anchor=east] at (7.9,0.6) {$\tilde{x} = 4$};
        \node[anchor=east] at (7.9,0.95) {$\tilde{x} = 4$};
        \node[anchor=east] at (7.9,1.3) {$\tilde{x} = 5$};
        \node[anchor=east] at (7.9,1.65) {$\tilde{x} = 5$};
    \end{tikzpicture}
    \caption{Agreement with: ``I finished the task fast.''.}
    \label{fig:efficiencyLikert}
    \vspace{-3mm}
\end{figure}
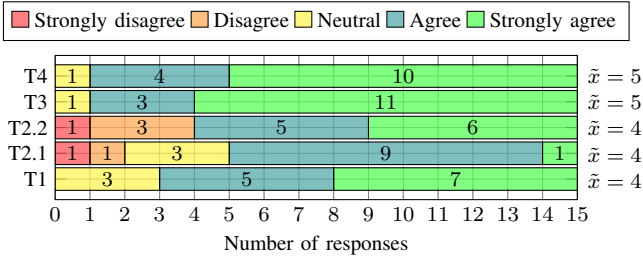

\figurename{~\ref{fig:efficiencyLikert}} shows the participants' subjective ratings of task completion time. Most participants agree with finishing quickly ($\tilde{x} = 4$). Task~1 shows mostly neutral to moderately positive agreement, while Tasks~2.1 and 2.2 had more varied but generally positive responses. Tasks~3 and 4 had strong agreement among participants, with most indicating they completed the tasks quickly, indicating a good efficiency.

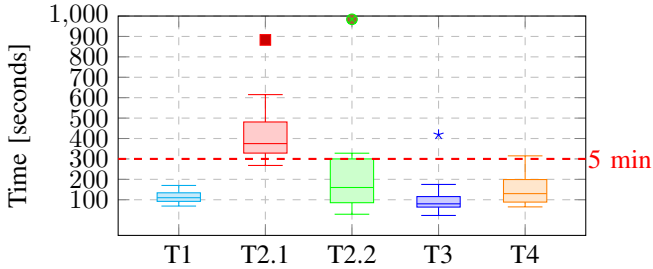
\begin{figure}[!h]
\vspace{-0.2cm}
\centering
\begin{tikzpicture}
\begin{axis}[
    boxplot/draw direction=y,
    xtick={1,2,3,4,5},
    xticklabels={T1, T2.1, T2.2, T3, T4},
    ylabel={Time [seconds]},
    ytick={100,200,...,1000},
    width=0.43\textwidth,
    height=4.5cm,
    grid=both,
    grid style={dashed,gray!50},
    boxplot/box extend=0.5,
    ymax=1000, 
    clip=false,
]
    \addplot+[boxplot,fill=cyan!20,draw=cyan] table[y index=0] {
        data
        132
        125
        73
        170
        88
        110
        136
        84
        96
        69
        122
        101
        152
        170
        109
    };
    \addplot+[boxplot,fill=red!20,draw=red] table[y index=0] {
        data
        375
        463
        414
        343
        268
        499
        327
        277
        330
        357
        403
        615
        571
        298
        883
    };
    \addplot+[boxplot,fill=green!20,draw=green] table[y index=0] {
        data
        29
        163
        328
        76
        58
        160
        160
        95
        73
        316
        300
        300
        142
        171
        984
    };
    \addplot+[boxplot,fill=blue!20,draw=blue] table[y index=0] {
        data
        136
        60
        68
        70
        28
        130
        175
        23
        86
        72
        80
        46
        101
        81
        419
    };
    \addplot+[boxplot,fill=orange!20,draw=orange] table[y index=0] {
        data
        188
        127
        91
        138
        71
        280
        247
        79
        92
        130
        137
        65
        315
        210
        86
    };
    \draw[red, thick,dashed] (axis cs:0.3,300) -- (axis cs:5.7,300);
    \node[anchor=west, red] at (axis cs:5.62,300) {5 min};
\end{axis}
\end{tikzpicture}
\caption{Task completion times.}
\label{fig:efficiencyBoxplot}
\vspace{-3mm}
\end{figure}

\figurename{~\ref{fig:efficiencyBoxplot}} presents the measured task completion times for each task (objective efficiency). The results show that participants took the least time on Task~1 and~3, indicating that these tasks could be solved quickly. Task~2.1 took the most time on average. The boxplots also reveal outliers, particularly for Tasks~2.1, 2.2, and 3, indicating that some participants required significantly more time than the average.

The task completion times were statistically analyzed per task. For this reason, we applied the Bonferroni-Holm correction to reduce the risk of Type I errors from repeated testing. As shown in \tablename{~\ref{tab:efficiencyStatistics}}, Task~1 shows a significant result ($t(14) = -22.2$, $p < 0.001$, Cohen's $d = 5.7$), indicating completion in under 5 minutes. Task~2.1 showed no significant difference ($Z = 3.25$, $p = 0.999$, $r = 0.8$), meaning it took more than 5 minutes. In contrast, Task~2.2 was completed in under 5 minutes ($Z = -2.03$, $p = 0.021$, $r = -0.6$). Task~3 also showed significance ($Z = -3.84$, $p < 0.001$, $r = -1$), as did Task~4 ($t(14) = -7.26$, $p < 0.001$, Cohen's $d = 1.9$). Therefore, the null hypothesis $H_{0, Efficiency}$ is rejected for 4 out of 5 tasks, providing evidence of good efficiency.

\begin{mdframed}
     \hypertarget{R2}{\textbf{Result 2: Efficiency.}} The participants took less than 5 minutes to complete Tasks 1, 2.2, 3, and 4. For Task 2.1, the participants took more than 5 minutes. 
\end{mdframed}
\vspace{-0.2cm}

\begin{table}[htbp]
    \setlength{\tabcolsep}{3pt} 
    \begin{scriptsize}
    \vspace{-2mm}
    \caption{Results of statistical tests regarding efficiency.}
    \begin{tabularx}{0.49\textwidth}{|l|c|c|c|c|c|c|X|}
        \hline
        \multicolumn{1}{|l|}{\multirow{2}{*}{}} & \multicolumn{2}{c|}{Shapiro-Wilk} & \multicolumn{2}{c|}{t-test} & \multicolumn{2}{c|}{Wilcoxon signed-rank} & \multicolumn{1}{c|}{\multirow{2}{*}{\pwertA}} \\
        \cline{2-7}
         & $W$ & $p$  & \twert & \pwert & \zwert & \pwert &  \\
        \hline\hline
        T1 & $0.96$ & $0.62$ & $-22.2$ & $<0.001$ & --- & --- & \multicolumn{1}{c|}{$<0,001$} \\
        \hline
        T2.1 & $0.83$ & $0.01$ & --- & --- & $3.25$ & $0.999$ & \multicolumn{1}{c|}{$0.999$} \\
        \hline
        T2.2 & $0.67$ & $<0.001$ & --- & --- & $-2.03$ & $0.02$ & \multicolumn{1}{c|}{$0.04$} \\
        \hline
        T3 & $0.07$ & $<0.001$ & --- & --- & $-3.84$ & $<0.001$ & \multicolumn{1}{c|}{$<0.001$} \\
        \hline
        T4 & $0.09$ & $0.05$ & $-7.3$ & $<0.001$ & --- & --- & \multicolumn{1}{c|}{$<0.001$} \\
        \hline
    \end{tabularx}
    \label{tab:efficiencyStatistics}
    \end{scriptsize}
    \vspace{-3mm}
\end{table}

\subsubsection{Satisfaction}
The satisfaction was measured using the short UEQ and resulted in a UEQ value of $1.93$. The results in \tablename{~\ref{tab:satisfactionStatistics}} show a significant difference between the measured satisfaction ($M = 1.9$, $SD = 0.6$) and the expected satisfaction ($M = 0.8$), ($t(14) = 7.5$, $p < 0.001$, Cohen’s $d = 1.9$). Therefore, the null hypothesis $H_{0, Satisfaction}$ is rejected, indicating that participants had a positive user experience with the dashboard (UEQ $> 0.8$).

\begin{mdframed}
    \hypertarget{R3}{\textbf{Result 3: Satisfaction.}} The participants had a positive user experience with the MVP (UEQ $> 0,8$).
\end{mdframed}
\vspace{-0.2cm}

\begin{table}[!htb]
    \vspace{-2mm}
    \footnotesize
    \caption{Results of statistical test regarding satisfaction.}
    \begin{tabularx}{0.47\textwidth}{|X|c|c|c|c|}
        \hline
        \multicolumn{1}{|l|}{\multirow{2}{*}{Satisfaction}} & \multicolumn{2}{c|}{Shapiro-Wilk} & \multicolumn{2}{c|}{t-test} \\
        \cline{2-5}
         & $W$ & $p$ & \twert & \pwert \\
        \hline\hline
        UEQ ratings & $0.90256$ & $0.10415$ & $7.4847$ & $<0.001$ \\
        \hline
    \end{tabularx}
    \label{tab:satisfactionStatistics}
    \vspace{-3mm}
\end{table}

\subsubsection{Team needs}\label{sec:team_needs}
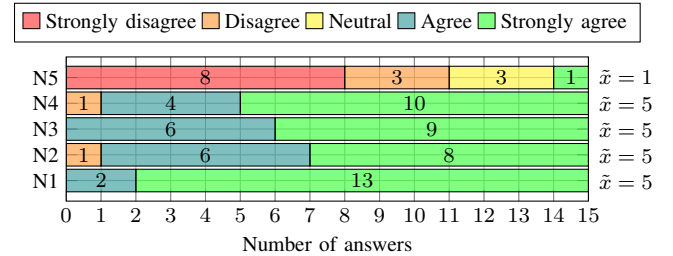
\begin{figure}[htbp]
    \centering
    \footnotesize
    \begin{tikzpicture}
        \begin{axis}[
            xbar stacked, 
            ytick=data, 
            yticklabels={N1, N2, N3, N4, N5}, 
            bar width=0.3cm,
            width=0.47\textwidth,
            height=3.5cm,
            xlabel={Number of answers}, 
            symbolic y coords={N1, N2, N3, N4, N5},
            nodes near coords, 
            legend style={at={(0.5,1.1)},anchor=south,legend columns=-1}, 
            xmin=0, xmax=15,
            xtick={0,1,...,15},
            enlarge y limits=0.2, 
            yticklabel style={align=center},
            grid=both,
        ]
            \addplot[fill=red,fill opacity=0.5, text opacity=1] coordinates {(0,N1) (0,N2) (0,N3) (0,N4) (8,N5)};
            \addplot[fill=orange,fill opacity=0.5, text opacity=1] coordinates {(0,N1) (1,N2) (0,N3) (1,N4) (3,N5)};
            \addplot[fill=yellow,fill opacity=0.5, text opacity=1] coordinates {(0,N1) (0,N2) (0,N3) (0,N4) (3,N5)};
            \addplot[fill=teal,fill opacity=0.5, text opacity=1] coordinates {(2,N1) (6,N2) (6,N3) (4,N4) (0,N5)};
            \addplot[fill=green,fill opacity=0.5, text opacity=1] coordinates {(13,N1) (8,N2) (9,N3) (10,N4) (1,N5)};
            \legend{Strongly disagree, Disagree, Neutral, Agree, Strongly agree}
        \end{axis}
        \node[anchor=east] at (7.9,0.25) {$\tilde{x} = 5$};
        \node[anchor=east] at (7.9,0.6) {$\tilde{x} = 5$};
        \node[anchor=east] at (7.9,0.95) {$\tilde{x} = 5$};
        \node[anchor=east] at (7.9,1.3) {$\tilde{x} = 5$};
        \node[anchor=east] at (7.9,1.65) {$\tilde{x} = 1$};
    \end{tikzpicture}
    \caption{Agreement with fulfillment of team needs.}
    \label{fig:needsLikert}
    \vspace{-2mm}
\end{figure}

One of the needs (cf.~\tablename{~\ref{tab:needs_of_team}}), N5 (tracking user activities), was intended as a control question and was expected to be rejected. The results (see \figurename{~\ref{fig:needsLikert}}) align with this expectation, with a median score of 1 and 8 ``strongly disagree'' responses. In contrast, the needs N1 to N4 were met, with median scores of 5, reflecting strong agreement. N1 (tracking visitor paths) received the highest support with 13 ``strongly agree'' and 2 ``agree'' responses. 
Similarly, N3 (tools for monitoring quality) followed closely with similar positive ratings. Needs N2 (marking inaccurate or incomplete resources) and N4 (overview of areas for improvement) had minimal neutral responses. These findings indicate successful implementation of N1 to N4 and the expected rejection of N5.

\section{Threats to Validity and Limitations}\label{sec:threats_limitations}
In the following, we discuss the threats to validity and limitations of both the action research and the experiment.

\subsection{Action Research}
A threat to the validity of the action research lies in the limited number of \href{https://doi.org/10.5281/zenodo.16908463}{\textit{\textcolor{link}{user stories}}} implemented. Although the initial set captured a broad range of insights into the C\&CB team's needs, only a subset was realized within the six-month period, and some had to be substantially adapted to meet practical constraints. In addition, several stories were better suited for direct integration into the ORKG frontend rather than a separate dashboard, as they mainly aimed at improving user experience in the service itself. To mitigate these risks, continuous communication and iterative feedback with the C\&CB team ensured that the implemented features still aligned with team needs.
Another key limitation is the generalizability of findings, as the insights of the action research are heavily tied to the specific context of the ORKG. While the second action research cycle aimed to enhance scalability and modularity, reusing SPARQL queries and adapting quality metrics across other (R)KGs remained challenging. 
The dashboard is not directly transferable as a whole to other (R)KG ecosystems due to varying graph structures and curation processes. Its overall design and components, such as visualizations, can be reused, but queries and metrics require customization. Fully generic dashboards are less effective than tailored ones, so we distill key takeaways for building customized curation dashboards.

\subsection{Experiment}\label{ch:experiment_threats}
The experiment evaluating the MVP faces several threats to validity. One concerns the design of the evaluation tasks, which, although simplified, may not fully capture the complexity of real-world scenarios, potentially affecting assessments of usability. To address this, tasks were developed and refined throughout the action research to reflect key curation and community-building activities. The C\&CB team did not know the tasks in advance, allowing seven members to participate without influencing results. 
Correlation analysis confirms no relationship between participants' prior involvement in the action research cycles and objective evaluation outcomes. Thus, there is no difference between results from C\&CB team members and the broader ORKG team, indicating that all participants can be treated as a single group. We still acknowledge a potential confirmation bias.

Additionally, the evaluation follows a single-case design, as neither a manual baseline nor comparison with alternative dashboards was feasible. The targeted curator workflows cannot realistically be completed without the MVP (except through disproportionate manual effort or coding/API queries), and, to the best of our knowledge, no comparable system exists that enables the retrieval of domain-specific metrics. This limitation reflects the \textit{fit-for-use} nature of (R)KG curation, where quality and effectiveness must be assessed within the specific context rather than against generic solutions. Consequently, the evaluation focused on the developed dashboard itself rather than relative performance.

Internal factors further complicate the experiment’s validity. Technical issues, such as small bugs or delays in the MVP, may have influenced participants' ability to complete tasks, thereby confounding the evaluation results. However, these challenges also provided valuable insights for improving the dashboard, guiding efforts to make it error-free and more user-friendly.
Variability in participant experiences, including differences in online versus in-person interactions, adds another layer of complexity. Such inconsistencies may have influenced task execution and feedback, potentially affecting the interpretation of the dashboard's performance.

Measuring objective effectiveness was complex, as there were multiple ways to achieve the desired goal in the interactive dashboard. Some participants chose a different way than originally intended, as it made more sense to them for the respective task. We considered a task to be completed successfully if the participant achieved the originally intended goal of the given task, regardless of the actual solution path.

\section{Discussion}\label{sec:discussion}
We discuss our results, share lessons learned, and present takeaways for researchers and human curators building dashboards for their (R)KGs, followed by future work.

\subsection{Results}
\hyperlink{R1}{Result 1: Effectiveness} shows that more than half of the tasks were completed successfully. Thus, we can consider the MVP as an effective solution. Measuring effectiveness was challenging, as some participants pursued different solutions paths than those originally intended. This behavior not only reflects the open-ended nature of real-world curation tasks but also suggests that the UI was intuitive enough to support context-aware decision-making. Tasks were considered successful when participants achieved the intended goal regardless of the actual solution path.

\hyperlink{R2}{Result 2: Efficiency} revealed that only one out of the five tasks (Task~2.1) could not be completed in under 5 minutes. Given the majority of the tasks were completed within the expected time, this result is a clear indicator that the MVP can be considered an efficient solution. The Task 2.1, focusing on analyzing visitor behavior, turned out to be the most complex, requiring high cognitive load due to its analytical depth and visually rich UI. Its complexity exceeded the assumptions made in the action research cycles, resulting in a higher effort to complete the task than expected.

In terms of subjective effectiveness and efficiency, the majority of participants agreed that the dashboard enabled them to complete their tasks successfully and quickly. However, some participants were unsure whether they used the intended solution path or completed the task as expected. This uncertainty illustrates the flexibility of the dashboard but also the need for clearer navigation.

\hyperlink{R3}{Result 3: Satisfaction} shows that participants had a positive user experience with the dashboard, indicating that the MVP is a satisfying solution. Participants emphasized its value for curation and community-building tasks, and most agreed that the team’s needs (cf. section~\ref{sec:team_needs}) were largely met. Members of the C\&CB team particularly noted how well the dashboard supports and enriches their work. One identified need could not be fulfilled due to internal technical constraints in the ORKG’s underlying structure rather than limitations of the developed solution.

Given these results, the MVP developed is a solid solution that supports the ORKG team, especially the C\&CB team, well in their curation and community-building processes for the ORKG regarding their needs, with a positive usability.

\subsection{Lessons Learned}
Based on our experiences during the development process, we offer the following reflections to support others aiming to build curation dashboards. Since (R)KG quality assessment is inherently domain-specific, these are no universal guidelines but rather practical insights that highlight key considerations and potential challenges.

Assessing the quality of an (R)KG involves multiple dimensions, including graph-structural metrics (e.g., connectivity), content-based metrics (e.g., completeness), metadata quality (e.g., provenance), and more. While some of these can be computed automatically, understanding the content quality in depth requires domain expertise. Through close collaboration with the C\&CB team and the use of action research, we were able to develop a dashboard tailored to their specific needs. Their input was crucial not only for interpreting metrics meaningfully but also for identifying which metrics were most relevant in the first place.

\begin{mdframed}
    \hypertarget{KT1}{\textbf{Takeaway 1: Talk to domain experts early.}}\\ The domain expert's insights are essential for identifying relevant metrics and interpreting them correctly.
\end{mdframed}
The C\&CB team’s deep understanding of the ORKG allowed them to contextualize metrics that might appear straightforward at first glance, such as the number of unique predicates. For example, a low number of predicates could indicate high reuse and adherence to a shared schema, which is typically interpreted positive. However, in the ORKG’s case, the flexible and user-driven modeling approach allows contributors to define their own predicates as needed. This leads to a lower reuse and more content creation with increased heterogeneity, as users are not bound by a strict vocabulary. 

\begin{mdframed}
    \hypertarget{KT2}{ \textbf{Takeaway 2: Prioritize tailored solutions.}}\\ Generic approaches fail to capture the specific characteristics of an (R)KG, so tailored solutions are required.
\end{mdframed}
In community-driven (R)KGs, naming conventions vary, and equivalent concepts may be modeled differently across (R)KGs. This variability complicates direct comparisons across (R)KGs and often requires explicit mapping between concepts. These challenges highlight the limits of fully automated or generic solutions for quality assessment and the importance of context-aware, expert-guided approaches. Especially for crowdsourced (R)KGs, such as the ORKG, understanding user behavior and platform dynamics becomes essential. By analyzing data from Matomo, we gained insights into how visitors interact with the platform, information that proved valuable not only for quality assessment but also for identifying areas for improving UI and navigation.

\begin{mdframed}
    \hypertarget{KT3}{\textbf{Takeaway 3: Crowdsourced (R)KGs require attention.}}\\
    Understanding user behavior and modeling diversity is crucial for quality assessment and UI improvements.
\end{mdframed}
While the dashboard was designed for modular extensibility and adaptation, the project revealed that meaningful cross-RKG comparisons are often constrained by fundamental structural and conceptual differences. In practice, the modular approach reaches its limits when deeper content-specific modeling and user behavior are considered. Quality assessment is then context-dependent, and attempts to generalize can risk overlooking domain nuances. This observation confirms a central lesson of this work: Tailored solutions, developed with domain experts, offer the most practical and effective way to curate knowledge of (R)KGs for quality improvement.

\begin{mdframed}
    \hypertarget{KT4}{\textbf{Takeaway 4: Modularity has boundaries.}}\\
    Structural and conceptual differences between (R)KGs limit how far components can be reused.
\end{mdframed}

\subsection{Future Work}

As the curation of (R)KGs remains a critical challenge, future work should explore how the dashboard can support scalable, hybrid curation workflows that combine human expertise with machine assistance: Automating routine tasks while keeping humans in control for contextual assessment and decision-making for modifying content inside the (R)KG. 

To better support human curators, especially those without technical background, an interactive interface, where curators define queries and directly generate visual analyses, making quality assessment more intuitive and responsive to evolving needs.
Furthermore, this research pointed to necessary adaptions in backend and data management infrastructure, not only to support data consumption and contribution, but also to sustain long-term, high-quality curation. Developers of (R)KG platforms should account for curator workflows as first-class requirements, not only those of end users. In this regard, action research proved essential for aligning the dashboard with real curation practices.

Future investigations across multiple (R)KGs may focus on identifying domain- and context-specific quality metrics, contributing to ongoing efforts in KG mapping and ontology alignment. Involving curators in this process can provide crucial insights for designing flexible, context-aware frameworks, although early work suggests that achieving such generalization that remains truly usable for in-depth assessment of individual (R)KGs is challenging.
The collaboration with the C\&CB team will continue as the dashboard evolves toward a fully tailored solution within the ORKG ecosystem.

\section{Conclusion}\label{sec:conclusion}
This research has demonstrated the value and feasibility of developing a tailored, modular dashboard for the curation of RKGs, specifically within the context of the ORKG. Through an iterative action research in close collaboration with the C\&CB team, we were able to design and implement \href{https://orkg.org/curation-dashboard}{\textit{\textcolor{link}{SciKGDash}}} as a minimum viable product (MVP) that reflects the C\&CB team’s needs and workflows. The experiment conducted with end users revealed promising results in terms of effectiveness, efficiency, and satisfaction, while also surfacing critical areas for improvement, particularly regarding UI complexity and performance optimization.

The development process highlighted the potential of the dashboard to support new curation workflows and foster community-building activities by making quality indicators more accessible and actionable. While the MVP marks an important milestone toward an extensible curation tool, it also exposed the inherent challenges in balancing generic applicability with specific requirements. The structural diversity of (R)KGs, such as the contrast between DBpedia and ORKG, underscores the need for continued research into flexible yet context-aware solutions.

Moreover, our findings reinforce the importance of involving end users throughout the development cycle. The close collaboration with the C\&CB team not only ensured the relevance of the implemented features but also surfaced deeper organizational and technical insights, such as the impact of dashboard design on internal ORKG structures and querying strategies. The study ultimately emphasizes that modular systems for (R)KGs must navigate a complex interplay between reuse, adaptability, and domain-specific constraints which is an endeavor that calls for both innovation and ongoing dialogue between developers and domain experts.


\section*{GenAI Usage Disclosure} ChatGPT was used to enhance the clarity and coherence of written text. All substantive content, analyses, and conclusions were produced entirely by the authors.

\bibliographystyle{ieeetr}
\bibliography{references}
\end{document}